\documentclass[twoside]{article}
\usepackage{fleqn,espcrc2}
\usepackage{graphicx}
\usepackage{amssymb}
\usepackage{float}

\newcommand{\AmS}{{\protect\the\textfont2
  A\kern-.1667em\lower.5ex\hbox{M}\kern-.125emS}}

\title{Four-Neutrino Scenarios\thanks{
Talk presented at
NOW 2000, Conca Specchiulla (Otranto, Italy), 9-16 Sep. 2000;
DFTT 47/00, hep-ph/0012236.}
}

\author{C. Giunti\thanks{
I would like to thank G. Mills and B. Louis
for useful information
on the LSND experiment.
}\\[0.2cm]
INFN, Sez. di Torino, and Dip. di Fisica Teorica,
Univ. di Torino, I--10125 Torino, Italy}
       
\begin{document}

\begin{abstract}
The main features of four-neutrino 3+1 and 2+2 mixing schemes
are reviewed,
after a discussion on the necessity of
at least four massive neutrinos
if the solar, atmospheric and LSND anomalies are
due to neutrino oscillations.
\end{abstract}

\maketitle

\begin{figure*}[t!]
\begin{center}
\includegraphics[bb=13 720 522 825,width=0.80\textwidth]{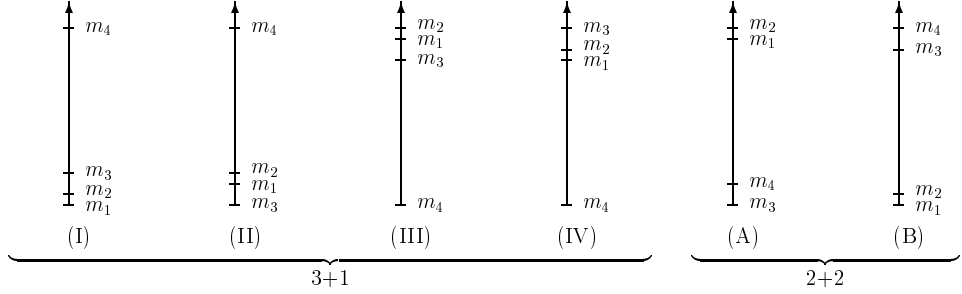}
\end{center}
\caption{ \label{4schemes}
Qualitative illustration of the possible four-neutrino schemes.
}
\end{figure*}

\section{Introduction}
\label{Introduction}

Solar and atmospheric neutrino experiments
have observed for a long time
anomalies that are commonly interpreted
as evidences in favor of neutrino oscillations
with mass squared differences
\begin{eqnarray}
&&
10^{-11} \, \mathrm{eV}^2
\lesssim
\Delta{m}^2_{\mathrm{SUN}}
\lesssim
10^{-4} \, \mathrm{eV}^2
\,,
\label{dm2sun}
\\
&&
10^{-3} \, \mathrm{eV}^2
\lesssim
\Delta{m}^2_{\mathrm{ATM}}
\lesssim
10^{-2} \, \mathrm{eV}^2
\,,
\label{dm2atm}
\end{eqnarray}
respectively
(see Refs.\cite{NOW2000-Smirnov,NOW2000-Kajita}).
More recently,
the accelerator LSND experiment has reported
the observation of
$\bar\nu_\mu\to\bar\nu_e$
and
$\nu_\mu\to\nu_e$
appearance~\cite{NOW2000-Spentzouris}
with a mass-squared difference
\begin{equation}
10^{-1} \, \mathrm{eV}^2
\lesssim
\Delta{m}^2_{\mathrm{LSND}}
\lesssim
10 \, \mathrm{eV}^2
\,.
\label{dm2LSND}
\end{equation}
The LSND evidence in favor of neutrino oscillations has not been confirmed
by other experiments,
but it has not been excluded either.
Awaiting an independent check of the LSND result,
that will probably come soon
from the MiniBooNE experiment~\cite{NOW2000-Spentzouris},
it is interesting to consider the possibility
that the results of solar, atmospheric and LSND experiments
are due to neutrino oscillations.
In this case,
the existence of the three mass-squared differences
(\ref{dm2sun})--(\ref{dm2LSND})
with different scales
implies that there are at least four massive neutrinos
(three massive neutrinos are not enough because
the three $\Delta{m}^2$'s
have different scales and
do not add up to zero).

Since the mass-squared differences
(\ref{dm2sun})--(\ref{dm2LSND})
have been obtained by analyzing separately the data
of each type of experiment
(solar, atmospheric and LSND)
in terms of two-neutrino mixing,
it is legitimate to ask if
three different mass squared are really necessary to fit the data.
The answer is ``yes'',
as explained in Section~\ref{Three}.

Although the precise measurement of the invisible width
of the $Z$ boson has determined that there are only three
active flavor neutrinos,
$\nu_e$, $\nu_\mu$, $\nu_\tau$,
the possible existence of at least four massive neutrinos is not a problem,
because in general flavor neutrinos are not mass eigenstates,
\textit{i.e.} there is \emph{neutrino mixing}
(see, \textit{e.g.}, Ref.\cite{BGG-review-98-brief}).

In general,
the left-handed component
$\nu_{\alpha L}$
of a flavor neutrino field
is a linear combination of the left-handed components
$\nu_{kL}$
of neutrino fields with masses $m_k$:
$
\nu_{\alpha L}
=
\sum_{k} U_{\alpha k} \nu_{kL}
$,
where $U$ is the unitary neutrino mixing matrix.
The number of massive neutrinos is only constrained
to be $\geq3$.
Following the old principle known as \emph{Occam razor},
we consider the simplest case of four massive neutrinos
that allows to explain all data
with neutrino oscillations~\cite{www-4nu}.
In this case,
in the flavor basis the usual three active neutrinos
$\nu_e$, $\nu_\mu$, $\nu_\tau$,
are associated with a sterile neutrino,
$\nu_s$,
that is a singlet of the electroweak group.

Taking into account the measured hierarchy
\begin{equation}
\Delta{m}^2_{\mathrm{SUN}}
\ll
\Delta{m}^2_{\mathrm{ATM}}
\ll
\Delta{m}^2_{\mathrm{LSND}}
\,,
\label{hierarchy}
\end{equation}
there are only six types of possible four-neutrino schemes,
which are shown in Fig.\ref{4schemes}.
These six schemes are divided in two classes:
3+1 and 2+2.
In both classes
there are two groups of neutrino masses
separated by the LSND gap, of the order of 1 eV,
such that the largest
mass-squared difference
generates the oscillations observed in the LSND experiment:
$\Delta{m}^2_{\mathrm{LSND}} = |\Delta{m}^2_{41}|$
(where
$\Delta{m}^2_{kj} \equiv m_k^2 - m_j^2$).
In 3+1
schemes there is a group of three neutrino masses
separated from an isolated mass
by the LSND gap.
In 2+2 schemes
there are two pairs of close masses
separated by the LSND gap.
The numbering of the mass eigenvalues in Fig.~\ref{4schemes}
is conveniently chosen in order to have always
solar neutrino oscillations generated by
$\Delta{m}^2_{21} = \Delta{m}^2_{\mathrm{SUN}}$.
In 3+1 schemes
atmospheric neutrino oscillations are generated by
$
|\Delta{m}^2_{31}|
\simeq
|\Delta{m}^2_{32}|
=
\Delta{m}^2_{\mathrm{ATM}}
$,
whereas in 2+2 schemes
they are generated by
$
|\Delta{m}^2_{43}|
=
\Delta{m}^2_{\mathrm{ATM}}
$.

In 1999
the 3+1 schemes were rather strongly disfavored by the experimental data,
with respect to the 2+2 schemes~\cite{BGGS-AB-99-brief}.
In June 2000 the LSND collaboration presented the
results of a new improved analysis of their data,
leading to an allowed region in the
$\sin^2 2\vartheta$--$\Delta{m}^2$ plane
($\vartheta$ is the two-generation mixing angle)
that is larger and shifted towards lower values of
$\sin^2 2\vartheta$,
with respect to the 1999 allowed region.
This implies that the 3+1 schemes are now marginally
compatible with the data.
Therefore,
in Section~\ref{3+1}
I discuss the 3+1 schemes,
that have been recently revived~\cite{Barger-Fate-2000-brief,%
Giunti-Laveder-ata-00,%
Peres-Smirnov-3+1-00}.
In Section~\ref{2+2}
I discuss the 2+2 schemes,
that are still favored by the data.

\section{Three $\Delta{m}^2$'s are necessary}
\label{Three}

Let us consider the general expression
of the probability of
$\nu_\alpha\to\nu_\beta$
transitions in vacuum valid for any number
of massive neutrinos:
\begin{equation}
P_{\nu_\alpha\to\nu_\beta}
=
\left|
\sum_k
U_{\alpha k}^* U_{\beta k}
\exp\left(-i\frac{\Delta{m}^2_{kj}L}{2E}\right)
\right|^2
\,,
\label{prob}
\end{equation}
where
$L$ is the source-detector distance,
$E$ is the neutrino energy,
and
$j$ is anyone of the mass eigenstate indices
(a phase common to all terms in the sum in Eq.(\ref{prob}) is irrelevant).

If all the phases
$\Delta{m}^2_{kj}L/2E$'s
are very small,
oscillations are not observable because
the probability reduces to
$P_{\nu_\alpha\to\nu_\beta} \simeq \delta_{\alpha\beta}$.
Since the LSND experiment has the smallest
average $L/E$,
of the order of $1 \, \mathrm{eV}^{-2}$,
at least one $\Delta{m}^2_{kj}$,
denoted by
$\Delta{m}^2_{\mathrm{LSND}}$,
must be larger than about
$10^{-1} \, \mathrm{eV}^2$
in order to generate the observed
$\bar\nu_\mu\to\bar\nu_e$
and
$\nu_\mu\to\nu_e$
LSND transitions,
whose measured probability
is of the order of $10^{-3}$.

Solar neutrino experiments observe
large transitions of $\nu_e$'s into other states,
with an average probability of about 1/2.
These transitions
cannot be generated by a
$\Delta{m}^2_{kj} \gtrsim 10^{-3} \, \mathrm{eV}^2$
because they should have been observed by the
long-baseline CHOOZ experiment~\cite{CHOOZ-99-brief}.
Hence,
at least another $\Delta{m}^2_{kj}$
smaller than about
$10^{-3} \, \mathrm{eV}^2$,
denoted by
$\Delta{m}^2_{\mathrm{SUN}}$,
is needed for the oscillations of solar neutrinos.

The necessary existence of at least a third $\Delta{m}^2_{kj}$
for atmospheric neutrino oscillations
is less obvious,
but can be understood by noticing that
a dependence of the transition probability
from the energy $E$
and/or from the distance $L$
is observable only if at least one phase
$\Delta{m}^2_{kj}L/2E$
is of order one.
Indeed, all the exponentials with phase
$\Delta{m}^2_{kj}L/2E \ll 1$
can be approximated to one,
whereas
all the exponentials with phase
$\Delta{m}^2_{kj}L/2E \gg 1$
are washed out by the averages
over the energy resolution of the detector
and the uncertainty in the source-detector distance.
Since the Super-Kamiokande atmospheric neutrino experiment
measures a variation of the oscillation probability
for
$L/E \sim 10^2 \div 10^3 \, \mathrm{eV}^{-2}$
(see Ref.\cite{NOW2000-Kajita}),
there must be at least one $\Delta{m}^2_{kj}$
in the range
$10^{-3} \div 10^{-2} \, \mathrm{eV}^2$,
which is out of the ranges allowed for
$\Delta{m}^2_{\mathrm{SUN}}$
and
$\Delta{m}^2_{\mathrm{LSND}}$.
Therefore,
at least a third $\Delta{m}^2_{kj}$,
denoted by
$\Delta{m}^2_{\mathrm{ATM}}$,
is needed for atmospheric neutrino oscillations.
This argument is supported by a detailed calculation
presented in Ref.\cite{Fogli-Lisi-Marrone-Scioscia-no3-99}.

\begin{figure}[b!]
\begin{center}
\includegraphics[bb=89 420 540 730,width=0.45\textwidth]{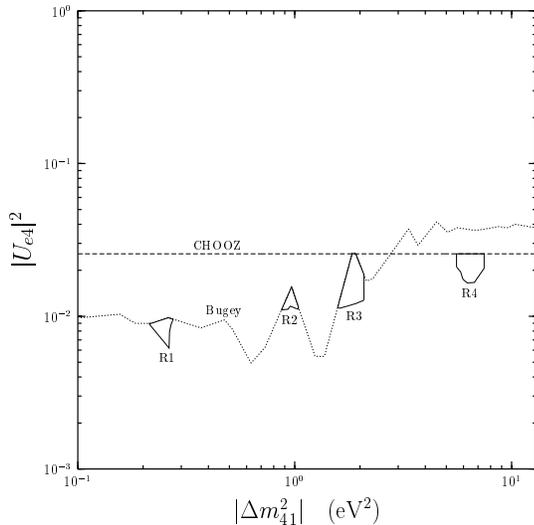}
\end{center}
\caption{ \label{uel4}
3+1 schemes.
Dotted and dashed lines:
$|U_{e4}|^2_{\mathrm{max}}$
from Bugey~\protect\cite{Bugey-brief}
and
CHOOZ~\protect\cite{CHOOZ-99-brief}.
Solid lines enclose the allowed regions.
}
\end{figure}

In the following sections we discuss some phenomenological aspects
of the four-neutrino schemes in Fig.\ref{4schemes},
in which there are three mass squared differences
with the hierarchy (\ref{hierarchy})
indicated by the data.

\begin{figure}[b!]
\begin{center}
\includegraphics[bb=89 420 540 730,width=0.45\textwidth]{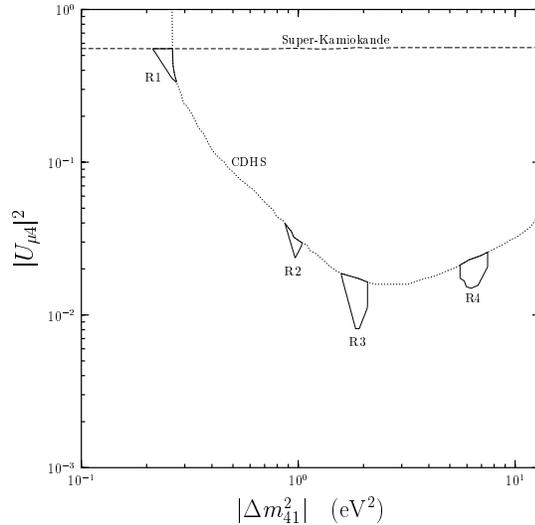}
\end{center}
\caption{ \label{umu4}
3+1 schemes.
Dotted and dashed lines:
$|U_{\mu4}|^2_{\mathrm{max}}$
from CDHS~\protect\cite{CDHS-brief}
and
Super-Kamiokande~\protect\cite{NOW2000-Kajita,BGGS-AB-99-brief}.
Solid lines:
allowed regions.
}
\end{figure}

\section{3+1 Schemes}
\label{3+1}

In 3+1 schemes
the amplitude of
$\nu_\alpha\to\nu_\beta$ and $\nu_\beta\to\nu_\alpha$
transitions in short-baseline neutrino oscillation experiments
(equivalent to the usual
$\sin^2 2\vartheta$
in the two-generation case)
is given by
(see, for example, Ref.\cite{BGG-review-98-brief})
\begin{equation}
A_{\alpha\beta}
=
A_{\beta\alpha}
=
4 |U_{\alpha4}|^2 |U_{\beta4}|^2
\,,
\label{14}
\end{equation}
and the oscillation amplitude
(again equivalent to the usual two-generation $\sin^2 2\vartheta$)
in short-baseline $\nu_\alpha$ disappearance experiments
is given by
\begin{equation}
B_\alpha
=
\sum_{\beta\neq\alpha} A_{\alpha\beta}
=
4 \, |U_{\alpha4}|^2 \left( 1 - |U_{\alpha4}|^2 \right)
\,.
\label{16}
\end{equation}

\begin{figure}[b!]
\begin{center}
\includegraphics[bb=89 420 540 730,width=0.45\textwidth]{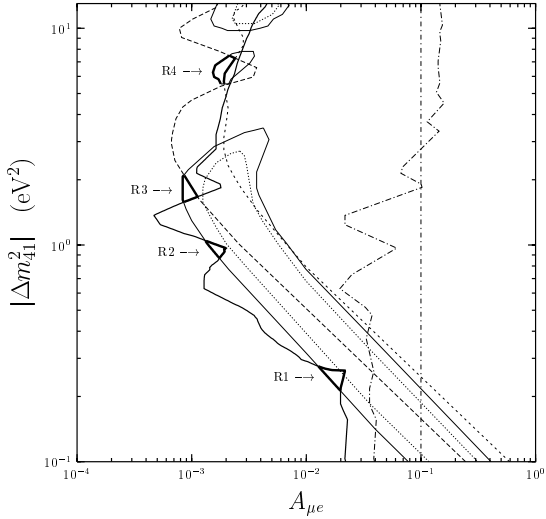}
\end{center}
\caption{ \label{31-amuel-allowed}
3+1 schemes.
Very thick solid line:
allowed regions.
Thick solid line:
disappearance bound (\ref{31bound}).
Dotted line:
LSND 2000 allowed regions at 90\% CL~\protect\cite{NOW2000-Spentzouris}.
Solid line:
LSND 2000 allowed regions at 99\% CL~\protect\cite{NOW2000-Spentzouris}.
Broken dash-dotted line:
Bugey exclusion curve at 90\% CL~\protect\cite{Bugey-brief}.
Vertical dash-dotted line:
CHOOZ exclusion curve at 90\% CL~\protect\cite{CHOOZ-99-brief}.
Long-dashed line:
KARMEN 2000 exclusion curve at 90\% CL~\protect\cite{NOW2000-Steidl}.
Short-dashed line:
BNL-E776 exclusion curve at 90\% CL~\protect\cite{BNL-E776}.
}
\end{figure}

\begin{figure}[b!]
\begin{center}
\includegraphics[bb=89 420 540 730,width=0.45\textwidth]{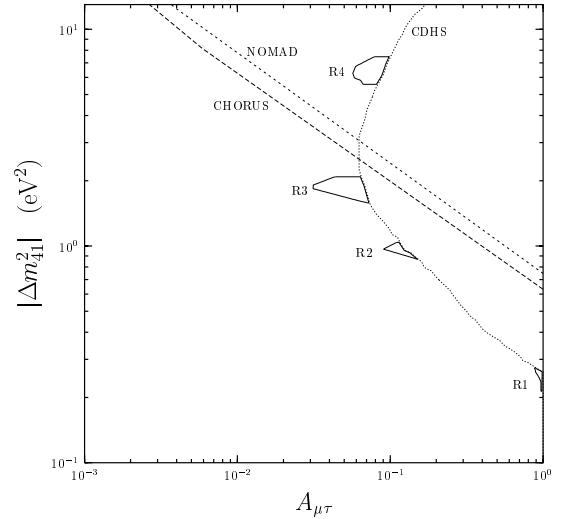}
\end{center}
\caption{ \label{amuta}
3+1 schemes with $|U_{s4}|^2 \ll 1$.
Solid lines enclose the allowed regions.
Long dashed line:
CHORUS exclusion curve at 90\% CL~\protect\cite{NOW2000-Zucchelli}.
Short dashed line:
NOMAD exclusion curve at 90\% CL~\protect\cite{NOW2000-Vidal-Sitjes}.
Dotted line:
CDHS exclusion curve at 90\% CL~\protect\cite{CDHS-brief}.
}
\end{figure}

Short-baseline $\bar\nu_e$ and $\nu_\mu$
disappearance experiments
put rather stringent limits
$B_e \leq B_e^{\mathrm{max}}$
and
$B_\mu \leq B_\mu^{\mathrm{max}}$
for
$|\Delta{m}^2_{41}|$
in the LSND-allowed region.
Taking into account also
the results of solar and atmospheric neutrino experiments,
Eq.(\ref{16}) implies that
$|U_{e4}|^2$
and
$|U_{\mu4}|^2$
are small
(see Ref.\cite{Giunti-Laveder-ata-00} and references therein):
\begin{equation}
|U_{e4}|^2 \leq |U_{e4}|^2_{\mathrm{max}}
\quad \mbox{and} \quad
|U_{\mu4}|^2 \leq |U_{\mu4}|^2_{\mathrm{max}}
\,,
\label{24}
\end{equation}
as shown by the dashed and dotted lines
in Figs.\ref{uel4} and \ref{umu4}.
These limits imply that
the amplitude
$A_{\mu e}$,
equivalent to the usual $\sin^2 2\vartheta$
in short-baseline
$\nu_\mu\to\nu_e$
and
$\bar\nu_\mu\to\bar\nu_e$
experiments,
is very small:
\begin{equation}
A_{{\mu}e}
\leq
4 |U_{\mu4}|^2_{\mathrm{max}} |U_{e4}|^2_{\mathrm{max}}
\,,
\label{31bound}
\end{equation}
so small to be at the border of compatibility
with the oscillations
observed in the LSND experiment.
Figure~\ref{31-amuel-allowed}
shows the comparison of the bound (\ref{31bound})
with
the LSND allowed region,
taking into account also the exclusion curves
exclusion curves of the KARMEN~\cite{NOW2000-Steidl}
and BNL-E776~\cite{BNL-E776} experiments.
One can see that there are four regions
that are marginally allowed,
denoted by R1, R2, R3, R4.

Let us denote by
$A_{{\mu}e}^{\mathrm{min}}$
the lower limit for $A_{{\mu}e}$
in the four allowed regions
in Fig.\ref{31-amuel-allowed}.
Then,
from
$A_{{\mu}e} = 4 |U_{\mu4}|^2 |U_{e4}|^2$
and the upper bounds (\ref{24}),
one can derive lower limits for
$|U_{e4}|^2$
and
$|U_{\mu4}|^2$:
\begin{equation}
|U_{e4}|^2
\geq
\frac{A_{{\mu}e}^{\mathrm{min}}}{4 |U_{\mu4}|^2_{\mathrm{max}}}
\,,
\quad
|U_{\mu4}|^2
\geq
\frac{A_{{\mu}e}^{\mathrm{min}}}{4 |U_{e4}|^2_{\mathrm{max}}}
\,.
\label{124}
\end{equation}
The upper and lower limits (\ref{24}) and (\ref{124})
for
$|U_{e4}|^2$ and $|U_{\mu4}|^2$
determine the allowed regions enclosed by solid lines
in Figs.\ref{uel4} and \ref{umu4}.

Summarizing the general properties of 3+1 schemes obtained so far,
from Fig.\ref{uel4}
we know that $|U_{e4}|^2$ is very small,
of the order of $10^{-2}$,
and from Fig.\ref{umu4}
we know that in the regions R2, R3, R4
$|U_{\mu4}|^2$ is also very small,
of the order of $10^{-2}$,
whereas in the region R1
$|U_{\mu4}|^2$
is relatively large,
$0.33 \lesssim |U_{\mu4}|^2 \lesssim 0.55$.
On the other hand,
the mixings of $\nu_\tau$ and $\nu_s$ with $\nu_4$
are unknown.

The authors of Ref.\cite{Barger-Fate-2000-brief}
considered the interesting possibility that
\begin{equation}
1 - |U_{s4}|^2 \ll 1
\,,
\label{fate}
\end{equation}
\textit{i.e.}
that the isolated neutrino $\nu_4$
practically coincides with $\nu_s$.
Notice, however,
that $|U_{s4}|^2$
cannot be exactly equal to one,
because LSND oscillations require that
$|U_{e4}|^2$ and $|U_{\mu4}|^2$
do not vanish,
as shown in Figs.\ref{uel4} and \ref{umu4},
and unitarity implies that
$1 - |U_{s4}|^2 \geq |U_{e4}|^2 + |U_{\mu4}|^2$.
The possibility (\ref{fate})
is attractive because
it represents a perturbation of the standard
three-neutrino mixing
in which a mass eigenstate is added,
that mixes mainly with the new sterile neutrino $\nu_s$
and very weakly with
the standard active neutrinos $\nu_e$, $\nu_\mu$, $\nu_\tau$.
In this case,
the usual phenomenology of three-neutrino mixing
in solar and atmospheric
neutrino oscillation experiments is practically unchanged:
the atmospheric neutrino anomaly would be explained by
dominant $\nu_\mu\to\nu_\tau$ transitions,
with possible sub-dominant $\nu_\mu\leftrightarrows\nu_e$
transitions constrained by the CHOOZ bound,
and
the solar neutrino problem would be explained by an
approximately equal mixture of
$\nu_e\to\nu_\mu$ and $\nu_e\to\nu_\tau$ transitions
(see, for example, Ref.\cite{BGG-review-98-brief}).
An appealing characteristic of this scenario
is the practical absence of transitions
of solar and atmospheric neutrinos into
sterile neutrinos,
that seems to be favored
by the latest data
(see \cite{NOW2000-Suzuki,NOW2000-Kajita,NOW2000-Ronga}).

\begin{figure}[b!]
\begin{center}
\includegraphics[bb=89 420 540 730,width=0.45\textwidth]{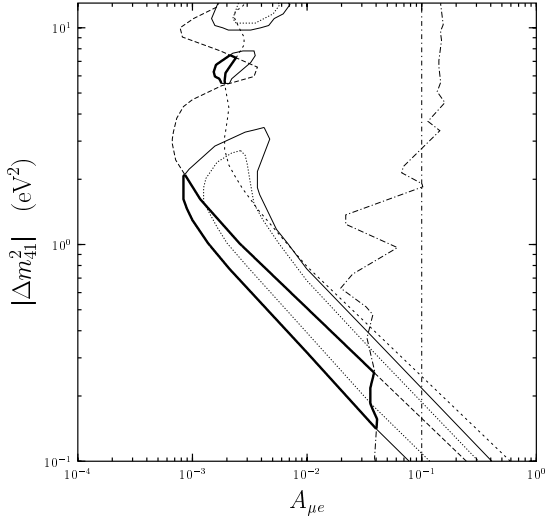}
\end{center}
\caption{ \label{22-amuel-allowed}
2+2 schemes.
See caption of Fig.\ref{31-amuel-allowed}.
}
\end{figure}

\begin{figure}[b!]
\begin{center}
\includegraphics[bb=89 420 540 730,width=0.45\textwidth]{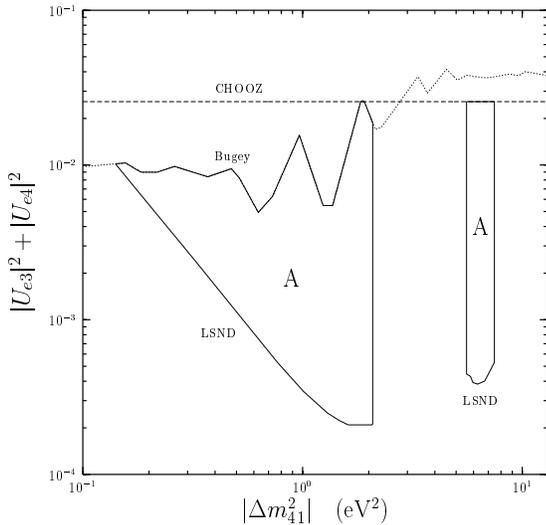}
\end{center}
\caption{ \label{sel}
2+2 schemes.
Dotted and dashed lines:
upper limit
from Bugey~\protect\cite{Bugey-brief}
and
CHOOZ~\protect\cite{CHOOZ-99-brief}.
The regions marked by ``A'' enclosed by solid lines are
allowed.
}
\end{figure}

\begin{figure}[b!]
\begin{center}
\includegraphics[bb=89 420 540 730,width=0.45\textwidth]{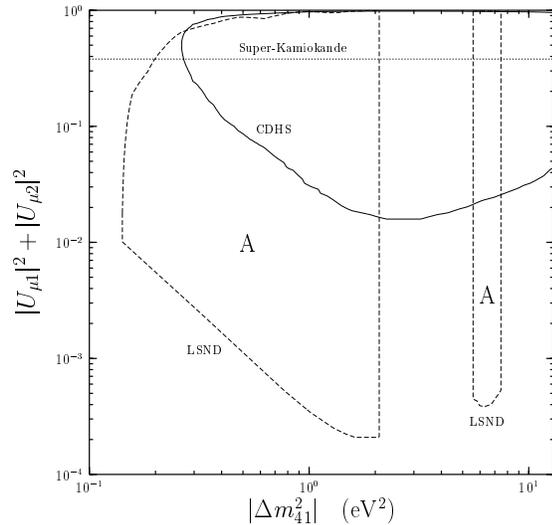}
\end{center}
\caption{ \label{smu}
2+2 schemes.
Solid, dotted and dashed lines:
limits
from CDHS~\protect\cite{CDHS-brief},
Super-Kamiokande~\protect\cite{NOW2000-Kajita,BGGS-AB-99-brief}
and
LSND~\protect\cite{NOW2000-Spentzouris,Giunti-JHEP-00-brief}.
The allowed regions are marked by ``A''.
}
\end{figure}

Another interesting possibility has been considered
in Ref.\cite{Giunti-Laveder-ata-00}:
\begin{equation}
|U_{s4}|^2 \ll 1
\,.
\label{large}
\end{equation}
This could be obtained,
for example,
in the hierarchical scheme I (see Fig.~\ref{4schemes})
with an appropriate symmetry keeping the
sterile neutrino very light,
\textit{i.e.}
mostly mixed with the lightest mass eigenstates.
Notice that nothing forbids
$|U_{s4}|^2$
to be even zero exactly.
The possibility (\ref{large})
is interesting because if it is realized there are relatively large
$\nu_\mu\to\nu_\tau$ and $\nu_e\to\nu_\tau$ transitions
in short-baseline neutrino oscillation experiments,
that could be observed in the near future.
This is due to the fact that
the unitarity of the mixing matrix implies that
$|U_{\tau4}|^2$
is large
($1-|U_{\tau4}|^2\ll1$ in the regions R2, R3, R4
and
$0.45 \lesssim |U_{\tau4}|^2 \lesssim 0.67$ in the region R1).
Therefore,
the amplitudes
$A_{\mu\tau} = 4 |U_{\mu4}|^2 |U_{\tau4}|^2$
and
$A_{e\tau} = 4 |U_{e4}|^2 |U_{\tau4}|^2$
of short-baseline
$\nu_\mu\to\nu_\tau$ and $\nu_e\to\nu_\tau$
oscillations
are suppressed only by the smallness of
$|U_{\mu4}|^2$ and $|U_{e4}|^2$
and lie just below the upper limits imposed by
the negative results of short-baseline
$\nu_\mu$ and $\bar\nu_e$
disappearance experiments.
Figure \ref{amuta}
shows the allowed regions
in the
$A_{\mu\tau}$--$|\Delta{m}^2_{41}|$
plane.
One can see that the region R4
is excluded by the negative results
of the CHORUS \cite{NOW2000-Zucchelli}
and NOMAD \cite{NOW2000-Vidal-Sitjes} experiments.
The other three regions are possible
and predict relatively large
oscillation amplitudes that could be observed
in the near future,
especially the two regions R2 and R3
in which
$A_{\mu\tau} \sim 4 \times 10^{-2} - 10^{-1}$.
An unattractive feature of this scenario
is its predictions of large $\nu_\mu\to\nu_s$ transitions
of atmospheric neutrinos,
that appear to be disfavored
by the latest data
(see \cite{NOW2000-Kajita,NOW2000-Ronga}).

\null \vspace{-1cm} \null

\section{2+2 Schemes}
\label{2+2}

The two 2+2 schemes in Fig.~\ref{4schemes}
are favored by the data because they do not suffer the
constraint imposed by the thick solid line in Fig.\ref{31-amuel-allowed},
that is valid only in 3+1 schemes.
Therefore,
all the part of the LSND region
in the $A_{\mu e}$--$\Delta{m}^2_{41}$ plane
that is not excluded
by other experiments is allowed,
as shown in Fig.\ref{22-amuel-allowed}.
For this reason,
the phenomenology of 2+2 schemes has been studied
in many articles~\cite{www-4nu}.

Figures~\ref{sel} and \ref{smu}
show the limits on the mixing of $\nu_e$ and $\nu_\mu$
obtained from the results of short-baseline,
solar and atmospheric experiments
\cite{www-4nu,Giunti-JHEP-00-brief}.
From Fig.~\ref{sel}
one can see that the mixing of $\nu_e$
with $\nu_3$ and $\nu_4$,
whose mass-squared difference
$\Delta{m}^2_{43}$
generates
atmospheric neutrino transitions,
is very small,
leading to a suppression of oscillations of $\nu_e$'s
in atmospheric and long-baseline experiments
\cite{BGG-bounds-98-brief}.

The mixing of $\nu_\tau$ and $\nu_s$
is almost unknown,
with weak limits obtained in recent fits
of solar
\cite{Concha-SNO-00}
and atmospheric data
\cite{Yasuda-fouratm-00,Fogli-Lisi-Marrone-fouratm-00-brief}.
For example,
it is possible that
both solar $\nu_e$'s and atmospheric $\nu_\mu$'s
oscillate into approximately equal mixtures of
$\nu_\tau$ and $\nu_s$'s.

In the future it may be possible to exclude
the scheme A
if it will be established with confidence
that the effective number of neutrinos
in Big-Bang Nucleosynthesis
is less that four.
In this case
$|U_{s3}|^2+|U_{s4}|^2 \ll 1$
\cite{Okada-Yasuda-97-brief}
and solar and atmospheric neutrino oscillations
occur, respectively,
through the decoupled channels
$\nu_e\to\nu_s$
and
$\nu_\mu\to\nu_\tau$.
It has been shown that in this scenario
the small mass splitting in scheme A between $\nu_3$ and $\nu_4$
is incompatible with radiative corrections
\cite{Ibarra-Navarro-4nu-00-brief}
and the effective Majorana mass in neutrinoless double-beta decay
in scheme A
is at the border of compatibility
with the experimental limit
\cite{Giunti-Neutrinoless-99-brief}.

\section{Conclusions}
\label{Conclusions}

Four-neutrino mixing is a realistic possibility
(if the solar, atmospheric and LSND anomalies are
due to neutrino oscillations).
It is rather complicated, but very interesting,
both for theory and experiments,
because:
it has a rich phenomenology;
the existence of a sterile neutrino is far
beyond the Standard Model,
hinting for exciting new physics;
there are several observable oscillation channels
in short-baseline and long-baseline experiments;
CP violation may be observable in
long-baseline experiments
\cite{www-4nu}.


\end{document}